# Noise-Limited Sensitivity in Cavity Optomechanical Molecular Sensing Enabled by Quantum Zero-Point Displacement Coupling and Strong Photon-Phonon Interaction for Chiral Detection

Giuseppina Simone

**Abstract**

This work demonstrates a quantum-limited optomechanical sensing platform for real-time detection and discrimination of chiral molecules, based on a multilayer hybrid plasmonic–mechanical resonator. By exploiting quantum zero-point motion and engineered photon–phonon interactions, the system achieves a displacement sensitivity on the order of $10^{-17}$ m/$\sqrt{\text{Hz}}$, approaching the standard quantum limit. The engineered multilayer stack, composed of alternating dielectric and metallic layers, supports high-Q mechanical resonances ($Q$ factors up to ~10,000) in the MHz regime, which coherently modulate the multilayer optical field *via* radiation pressure and dynamical backaction. Power spectral density measurements exhibit distinct mechanical peaks at 0.68, 2.9, 4.3, 5.5, and 6.8 MHz, with single-photon coupling rates up to 2.6 times the intrinsic baseline, enabling strong optomechanical transduction. Lorentzian fitting confirms narrow mechanical linewidths, while the total noise floor—including thermal, shot, and technical contributions—remains below $10^{-16}$ N/$\sqrt{\text{Hz}}$, ensuring robust detection of sub-piconewton forces. Time-resolved Raman spectroscopy, typically insensitive to chirality, here reveals enantioselective dynamics due to asymmetric optomechanical coupling in the hybrid cavity, enabling clear spectral differentiation of D- and L-enantiomers. Finite-element simulations corroborate the strong spatial overlap between optical confinement and mechanical displacement fields. This study establishes a scalable and tunable platform for high-sensitivity, quantum-limited detection of chiral molecules, with potential applications in coherent control, precision molecular spectroscopy, and ultrasensitive chemical sensing.

Optomechanics ,Plasmonic resonator ,Chiral detection ,Enantioselective sensing ,Quantum-limited measurement ,Zero-point fluctuations ,Noise spectral analysis

# Introduction

Photon-phonon coupling lies at the heart of interactions enabling mutual control and signal transduction between optical and mechanical domains, with profound implications for optical communication, quantum information processing, and sensing applications (1; 2; 3). The photon-phonon interaction can be enhanced by engineering the electromagnetic field or by controlling quantum pathways through resonant excitation (4; 5). A particularly powerful approach confines both photons and phonons within nanoscale cavities, where spatially coherent coupling selectively enhances phonon resonances (6; 7; 8).

Raman scattering remains a versatile and widely used technique to probe phonons and electron-phonon interactions, providing insight into vibrational and electronic properties at both bulk and nanoscale (9; 10; 11). In quantum Raman processes, incident photons excite virtual electronic states that mediate phonon creation or annihilation, producing inelastically scattered photons shifted by characteristic phonon energies (12; 13). These quantum pathways, determined by the electronic band structure and vibrational modes, critically govern the photon-phonon interaction strength (14; 15).

Conventional Raman theory typically assumes the dipole approximation, valid when photon wavelengths far exceed the spatial extent of phonon displacement fields (9; 16). Under the dipole approximation, selection rules follow from symmetry and wavevector conservation. However, in nanostructured materials or for high-momentum phonons, where optical wavelengths become comparable to phonon spatial scales, the dipole approximation fails and conventional selection rules may break down (17; 18; 19). Furthermore, the commonly employed bond polarizability model neglects photon wavevector dependence in matrix elements, an assumption increasingly challenged by spatial dispersion and near-field effects in confined geometries (20; 21; 22).

Engineering spatial overlap between photons and phonons in bulk materials at sub-wavelength scales remains technically demanding (23; 24). In contrast, van der Waals layered materials offer an inherently tunable platform where phonon confinement along the out-of-plane direction generates standing waves that enable momentum and wavelength matching with optical fields (25; 26; 27; 28; 29; 30). By adjusting layer thickness and dielectric surroundings, one can precisely control photon-phonon coupling beyond bulk limitations (31; 32). The multilayer structures also act as Fabry–Pérot cavities, creating spatially modulated optical fields that enhance local light–matter interaction and modulate coupling strength (33; 34; 35; 36).

Interference effects within multilayers modify both spatial and spectral optical mode profiles, giving rise to altered or novel Raman selection rules shaped by symmetry, momentum conservation, and mode overlap (37; 38; 39; 40; 41). In this work, we exploit the unique optomechanical characteristics of layered semiconductors to enhance Raman sensitivity to subtle vibrational features, particularly those relevant to chiral molecular detection. By tailoring the multilayer structure for optomechanical resonance, we show that the system can selectively amplify chiral vibrational responses, enabling dynamic detection of l- and d-enantiomers via their differential interaction with confined cavity fields. The resulting platform provides a quantum-limited approach to real-time chiral sensing, powered by enhanced photon–phonon coupling in a hybrid photonic–mechanical cavity.

We demonstrate that specific multilayer thicknesses act as intrinsic optical and phonon cavities supporting spatially varying electromagnetic fields and quantized layer-breathing phonon modes (42; 43; 44; 45; 46). When photon and phonon wavevectors align, photon-phonon coupling is selectively enhanced, enabling Raman activation of modes typically forbidden by bulk selection rules (14; 32; 28; 44; 36).

The spatially modulated optomechanical interactions enable the sensitive detection of molecular enantiomers via their distinct Raman fingerprints. By harnessing multilayer-induced interference and symmetry breaking within multilayers, we establish a platform for optomechanical sensing with high specificity and enhanced signal-to-noise. The Raman scattering intensity, governed by electron-photon and electron-phonon matrix elements, is thus modulated by spatial overlap and spectral detuning within the multilayer cavity (5; 42; 38; 40). Our findings open new pathways for exploiting optomechanical phenomena in layered materials toward advanced chiral sensing and molecular identification.

# Experimental Methods

## Multilayer Fabrication

A multilayer photonic structure composed of alternating dielectric and noble metal layers was fabricated for optical and Raman spectroscopic studies (47; 37; 48). The layer sequence, thicknesses, and material types are listed in **Table 1**, and the corresponding schematic representation is shown in Figure 1a. Fabrication began with a 4-inch, double-side polished, (100)-oriented p-type silicon wafer, which served as the mechanical support. The wafer was etched in order to achieve a thickness of $30 nm$ A base dielectric layer of silicon dioxide ($SiO_2$) was deposited using a standard plasma-enhanced chemical vapor deposition process, optimized for film uniformity and minimal surface roughness. A 40 nm silver (Ag) layer was deposited *via* electron beam evaporation at a base pressure below $5 \times 10^{-6}$ Torr to ensure high film purity and conformal coverage. The Ag layer covered the patterned grating and background substrate regions. Subsequent to this, a bilayer metal stack consisting of 10 nm of titanium as an adhesion layer and 50 nm of platinum (Pt) as the primary structural metal was deposited using radiofrequency sputtering under argon plasma at 3 mTorr (49,50).

*Layer structure, thickness, and refractive index (n) at $\lambda = 650$ nm for the fabricated multilayer.*

| Material Layer | Thickness (nm) | Refractive Index $n$ at 650 nm |
|---|---|---|
| Silver (Ag) | 30 | $0.135 + 3.999i$ |
| Silicon Dioxide ($SiO_2$, front) | 300 | 1.46 |
| Silicon (Si, substrate) | – (bulk) | $3.88 + 0.018i$ |
| Silicon Dioxide ($SiO_2$, back) | $3 \times 10^4$ | 1.46 |
| Titanium (Ti, adhesion) | 10 | $2.7 + 3.4i$ |
| Platinum (Pt) | 50 | $2.65 + 5.10i$ |

## Optical Measurements

The optical properties of the fabricated structure were characterized using a custom-built Kretschmann configuration setup (51). The system was illuminated by a tunable continuous-wave laser centered at 650nm. The beam was first passed through a fiber collimator (Thorlabs RC12FC-P01) to produce a well-collimated output and then directed through a double GlanTaylor calcite polarizer to achieve precise p-polarization alignment. An iris diaphragm (Thorlabs D25SZ) was inserted into the beam path to spatially filter stray light and minimize background noise. The sample was optically coupled to the base of a BK7 right-angle prism using index-matching oil to ensure efficient excitation of surface plasmon polaritons (52). After interaction with the multilayer structure, the reflected beam passed through a secondary polarizer (optimized for 650 − 1000 nm operation) for polarization filtering and was then detected by a silicon photodiode with a spectral response bandwidth extending up to 960 nm (53). To suppress electrical noise and improve signal stability, a ceramic disk capacitor was connected in parallel with the detection circuit. For Raman spectroscopy, the detection module was positioned in front of the prism. A silicon photodiode detector (wavelength range: 340 − 1100 nm) captured the inelastically scattered light. The Raman signal was analyzed using a PicoTech Series 5000 spectrum analyzer. A CMOS camera recorded the spatially resolved Raman spectra. Penicillamine was selected as the analyte, deposited in powder form (1 µg per droplet) following a solvent-free sample preparation protocol. Each measurement was repeated three times to confirm reproducibility and statistical significance.

.

## Numerical Analysis

Finite element simulations were carried out using MATLAB's Partial Differential Equation Toolbox (Edition 2024). A 3D mesh of the multilayer structure was generated based on the fabricated geometry. Modal analysis was performed under mechanical boundary conditions that imposed a uniaxial compressive stress of 50 MPa on opposing mirror surfaces, mimicking plasmonic-mechanical interaction scenarios. The output included eigenfrequencies, stress distributions, and displacement fields. Electromagnetic simulations were conducted using the transfer matrix method, which relates the complex amplitudes of incident, reflected, and transmitted electric fields. The simulation parameters included angle of incidence ($\vartheta$) and wavelength ($\lambda$) sweeps over experimentally relevant ranges. Optical constants for all materials were taken from well-established references: Ag, Pt, Si, and $SiO_2$ values were sourced from Palik's Handbook of Optical Constants of Solids; fused silica data was obtained from the WVASE software database; and the refractive index of NBK7 prism glass was taken from Schott's official material data sheets. The combined mechanical and optical simulation results were used to interpret experimental trends and validate observed dispersion features in the fabricated multilayer stack. To analyze the optomechanical response and the chiral-specific Raman features of L- and D- enantiomorph, we implemented a systematic processing routine that combines spatial, spectral, and temporal resolution of the measured data stack. Each acquired dataset consists of a 3D intensity matrix where the two in-plane dimensions represent the spatially resolved Raman spectrum and the third dimension corresponds to the time evolution under optical and mechanical excitation. The raw stack is pre-processed by identifying the most intense region along the spatial dimension (toprow and bottomrow), effectively isolating the portion of the optomechanical cavity where the photon–phonon interaction is strongest. For each selected spectral window (Imagex), the matrix is reduced by averaging the intensity along the high-signal region to improve the signal-to-noise ratio and to extract meaningful spectral shifts. The resulting subset is then visualized as a time versus Raman shift map, revealing how the vibrational modes evolve as a function of interaction time with the optical field. To better interpret transient effects and local intensity fluctuations, a lateral integrated profile is computed for each time frame by summing or averaging the spectral intensity along the Raman shift axis, producing a complementary time-resolved trace of the total scattered signal. The data are further interpolated to smooth residual artifacts and to highlight subtle dynamical trends. The dual representation, a 2D colormap of spectral intensity and a side profile of integrated signal, provides a robust visualization of how chiral molecules interact differently with the optomechanical resonator. The L- and D- enantiomers exhibit distinguishable temporal patterns and spectral shifts that reflect their asymmetric interaction with the local electromagnetic fields and mechanical displacement. The approach leverages the hybrid plasmonic–mechanical cavity's ability to amplify photon–phonon coupling beyond the standard Raman selection rules, enabling the discrimination of enantiomers by their dynamic vibrational response. The extracted time–wavenumber maps, combined with the quantitative side profiles, thus serve as a powerful tool for real-time chiral analysis using quantum-limited optomechanical detection.

## Results and Discussion

### Plasmonic resonance mediated optomechanical coupling: physics of the system and an experimental characterization

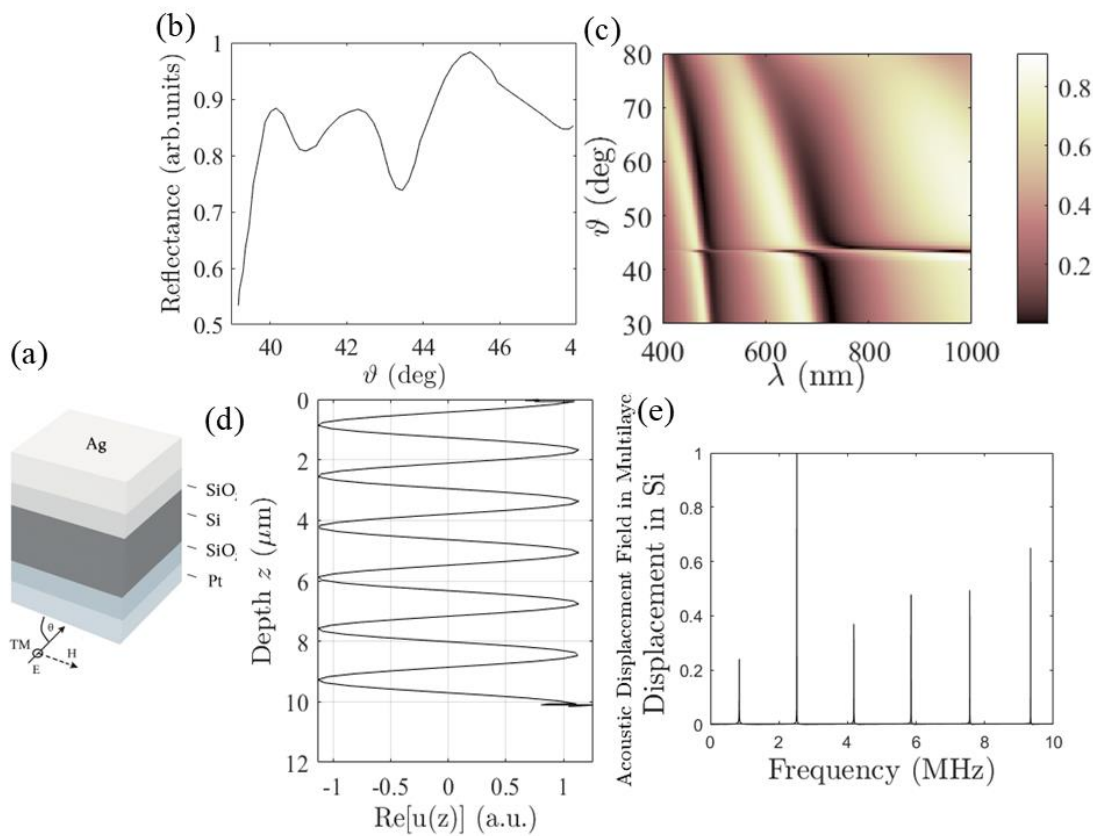

*Figure 1. Architecture and characterization of the hybrid optomechanical sensor. (a) Schematic of the five-layer structure on a prism enabling optical excitation via the Kretschmann configuration. (b) Angle-resolved experimental reflectance spectrum showing plasmonic and Bloch dips at 43.5° and 41.8°, respectively. (c) Reflectance map illustrating mode hybridization and anti-crossing behavior. (d) Simulated mechanical displacement highlighting acoustic*

*confinement in the Si layer. (e) Mechanical displacement spectrum with a pronounced resonance at* 2.8 MHz.

The hybrid optomechanical sensor presented in this work is designed to combine both plasmonic and mechanical resonances to achieve enhanced transduction sensitivity. The core structure consists of a five-layer heterostructure (Pt/SiO$_2$/Si/SiO$_2$/Ag) carefully engineered to simultaneously support confined optical fields and high-$Q$ mechanical vibrations (**Figure 1a**).

For the plasmonic analysis, the multilayer is positioned onto a high-refractive-index prism, enabling excitation *via* the Kretschmann configuration. A transverse magnetic polarized continuous-wave laser is introduced through the prism at variable incidence angles $\vartheta$, allowing evanescent coupling into the multilayer. The optical response of the device is characterized by angular reflectance spectroscopy, which reveals a distinct dip in reflectance at $\vartheta \approx 43.5°$, corresponding to the resonant excitation of a surface plasmon polariton mode at the SiO$_2$/Ag interface (**Figure 1b**). However, a Bloch mode is also visible at $\vartheta \approx 41.8°$. The plasmonic dispersion landscape of the multilayer is illustrated by a reflectance map as a function of wavelength $\lambda$ and incidence angle $\vartheta$ (**Figure 1c**). The observed anti-crossing behavior demonstrates strong coupling between the optical multilayer modes and the plasmon polariton branch, resulting in hybrid plasmon-Bloch states. The modes are crucial for the sensor's operation, as their spectral position and intensity are sensitive to sub-nanometer mechanical displacements of the layered structure. According to the dispersion plot, the plasmonic resonance occurs at $\omega_{\text{spp}} = 6.38 \times 10^{14}$ Hz with a linewidth of $\kappa = 5 \times 10^{12}$ Hz, yielding an optical quality factor $Q = 128$. The narrow spectral feature provides strong field enhancement at the metal-dielectric boundary, which is essential for sensitive displacement detection in optomechanical systems.

Mechanical properties are intrinsically linked to the sensor's optical modulation mechanism. The proposed design features a central SiO$_2$/Si/SiO$_2$ region that acts as an acoustic multilayer, supporting standing longitudinal modes primarily localized within the Si layer due to acoustic impedance mismatch with the surrounding oxide and metal layers. In fact, finite-element simulations of the normalized displacement ($\text{Re}(u(z))$) field show that the acoustic energy is concentrated along the vertical axis of the structure, with minimal penetration into the Ag and Pt layers (**Figure 1d**). The mechanical confinement enables efficient modulation of the optical path length in response to thermal or driven excitations. The displacement spectrum extracted from frequency-resolved mechanical simulations (**Figure 1e**) reveals a pronounced resonance peak at the fundamental frequency $\Omega_m = 2\pi \times 2.8$ MHz. The mode is expected to correspond to a longitudinal standing wave involving both in-plane and out-of-plane deformations within the Si layer. The mechanical mode exhibits a high displacement amplitude and narrow linewidth, facilitating enhanced optomechanical interaction through modulation of the multilayer's optical boundary conditions. The mechanical mode at frequency $\Omega_m = 2\pi \times 2.8$ MHz features a well-defined spectral peak in the simulated displacement spectrum. On the other hand, the peak displacement amplitude, $x = 0.34$ nm, is obtained by evaluating the out-of-plane displacement field under a unit-amplitude harmonic excitation at $\Omega_m$. The excitation is applied solely to extract the mode shape and associated field distributions, and does not represent a physically applied load in the system. Together with an effective motional mass of $m_{\text{eff}} = 9.9$ ng and a mechanical damping rate of $\gamma_m = 280$ Hz, the quality factor of the resonance is calculated as $Q_m = \Omega_m/\gamma_m \approx 1.0 \times 10^4$, confirming strong mechanical confinement and low energy dissipation. The vibrational energy stored in the mode, $U = 1/2\, m_{\text{eff}}\Omega_m^2 x^2 \approx 4.4 \times 10^{-18}$ J, indicates that the multilayer structure can sustain high-amplitude mechanical oscillations at resonance under minimal driving conditions. It is worth noting that, for enhancing optomechanical coupling in the hybrid sensor architecture, the well-resolved nature of the mode and its sensitivity to boundary motion are fundamental.

The sensing strategy here exploited is based on the combined optical and mechanical response of the multilayer structure, which forms the foundation of the device's operation. The key mechanism involves the coupling between localized plasmonic fields and mechanical displacement, so that, even small structural deformations shift the optical resonance, enabling external forces to be detected through changes in reflectance.

The photonic-phononic coupling in the multilayer system arises from dispersive optomechanical interactions, governed by the intrinsic optical and mechanical characteristics of the structure. In the optomechanical regime, coupling between confined optical and mechanical modes is influenced by both classical and quantum forces. Excitation of plasmonic resonances by a laser induces radiation pressure on the internal boundaries, resulting from photon momentum transfer. The radiation pressure acts as a distributed mechanical load, modeled as tensile stress applied to the inner surfaces of the top and bottom metallic layers, analogous to classical static tension.

To characterize the mechanical behavior under optomechanical loading, a steady-state tensile stress of approximately 50 MPa was applied to the inner faces of the top and bottom layers in the numerical simulations.

The resulting structural response is summarized in **Figure 2a**. Panels (i–ii), which shows the frequency-domain spectra of representative mechanical resonances, highlighting distinct modal behaviors. The spectral analysis identifies two fundamental mechanical resonances with distinct characteristics: the lower-frequency mode at 2.1 MHz corresponds to significant out-of-plane bending motion, while the higher-frequency mode at 3.4 MHz is dominated by purely in-plane elastic dynamics, including twisting and lateral flexure components. The out-of-plane mode exhibits a quality factor of approximately $8 \times 10^3$ with clear signs of nonlinear stiffening under increased stress, whereas the in-plane mode achieves a higher quality factor exceeding $1.2 \times 10^4$, indicative of more efficient energy storage in layer-parallel vibrations. The robust separation of deformation modes is critical for understanding and optimizing both optomechanical coupling strength and overall sensor performance. Panels (iii) and (iv) in Figure 2a present the time-domain displacements projected along the in-plane $X$ and $Y$ directions, revealing the temporal evolution of the elastic vibrations. The deformation profiles are smooth and periodic, capturing the clean harmonic oscillations of the structure along both axes, which is consistent with a high-quality elastic response. The traces not only confirm the high-$Q$ behavior ($Q > 10^4$) but also reveal subtle differences in the vibrational patterns: the $X$-projection displays the characteristic signature of torsional motion, while the $Y$-projection shows the expected profile of axial flexure. Such coherent in-plane motion is particularly important for efficient optomechanical coupling, as it enhances overlap with the optical field and enables dispersive or strain-mediated modulation of optical properties. The spatial deformation of the structure under stress, revealing distinct bending along the $x$-axis and twisting along the $z$-axis (**Figure 2b**). These mode shapes highlight the fundamentally different mechanical behaviors: in-plane bending remains mostly elastic and uniform, while out-of-plane twisting introduces geometric nonlinearities that can strongly affect frequency stability. The difference is further illustrated in the noise spectrum (**Figure 2c**), where the thermomechanical noise floor reaches $1 \times 10^{-17}$ m$^{-1}$, approaching the fundamental quantum limit.

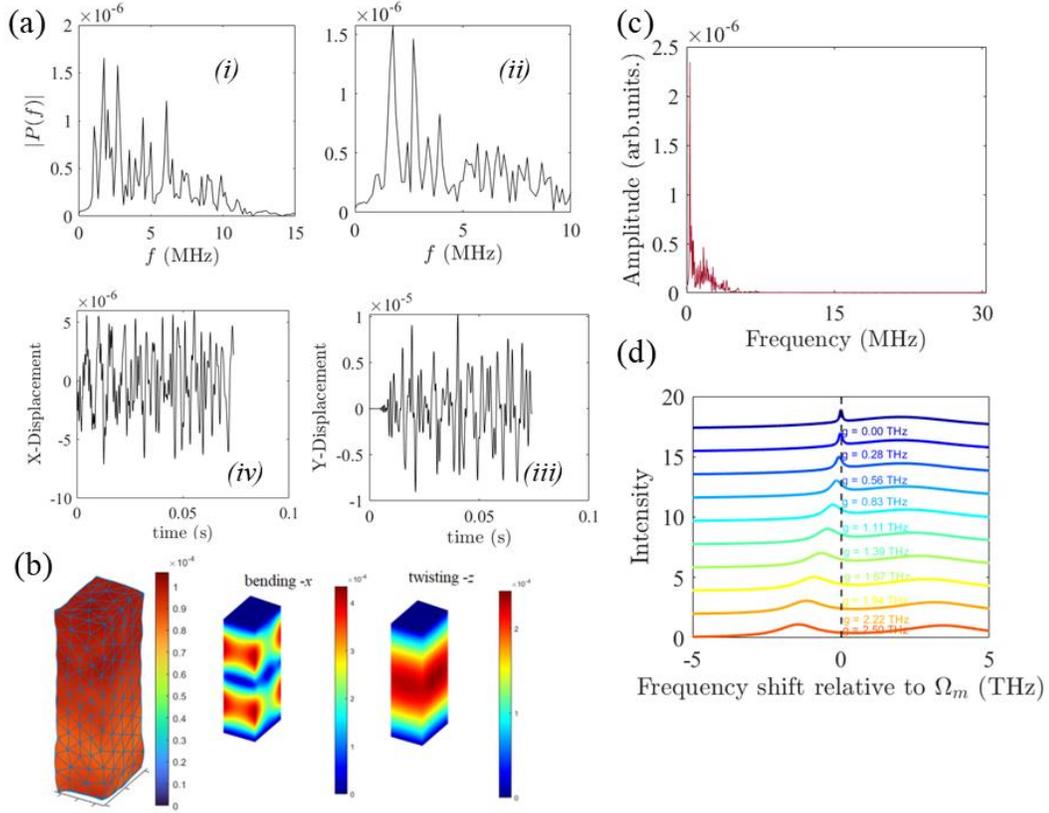

*Figure 2. Optomechanical characterization. (a) Frequency-domain and time-domain mechanical response of the multilayer resonator under axial loading. (i–ii) Frequency-domain spectra of representative mechanical resonances under 50 MPa quasi-static uniaxial stress, showing two fundamental modes at approximately 2.1 MHz (out-of-plane bending) and 3.4 MHz (in-plane twisting/flexure). (iii–iv) Time-domain displacement fields projected along the in-plane X and Y directions, illustrating the temporal evolution and spatial profiles of the elastic vibrations. Units of displacement are normalized to the zero-point fluctuation amplitude where indicated. (b) Simulated structural response under static radiation pressure loading on the inner surfaces of the metallic layers, showing combined out-of-plane bending and in-plane deformation modes. Units of geometry in µm. (c) Thermomechanical noise spectrum approaching the quantum regime. The noise floor reaches approximately $1 \times 10^{-17}$ m$^{-1}$. (d) Strain-induced shift of the cavity resonance frequency under axial loading. The modified optical path length of the multilayer cavity produces a measurable blue-shift in the plasmonic resonance that directly reflects the underlying mechanical deformation.*

Mechanical deformation modifies the optical path length within the multilayer structure, shifting the photonic resonance frequency and thereby enabling modulation of the optical response. The mechanical oscillator exhibits intrinsic quantum fluctuations characterized by the zero-point amplitude $x_{\text{zpf}}$, which defines the characteristic scale of Brownian motion in the quantum ground state and sets the fundamental displacement limit for quantum-limited mechanical measurements. For a mechanical resonance frequency of $\Omega_m = 2\pi \times 2.8$ MHz and an effective mass $m_{\text{eff}} = 9.9$ ng, the zero-point displacement is calculated as $x_{\text{zpf}} = \sqrt{\frac{\hbar}{2m_{\text{eff}}\Omega_m}} \approx 5.5 \times 10^{-16}$ m. In comparison, the classical steady-state displacement due to radiation pressure, under an optical power of $P_{\text{cav}} = 1.6\ \mu$W, is given by $x_{\text{cl}} = \frac{F_{\text{rad}}}{m_{\text{eff}}\Omega_m^2} = \frac{2P_{\text{cav}}/c}{m_{\text{eff}}\Omega_m^2} \approx 3.5 \times 10^{-18}$ m where $F_{\text{rad}} = \frac{2P_{\text{cav}}}{c}$ is the radiation pressure force acting on the structure. The comparison between $x_{\text{zpf}}$ and $x_{\text{cl}}$ is fundamental for distinguishing quantum-limited behavior from classical motion. In the multilayer configuration under study (see **Table 2**), the inequality $x_{\text{zpf}} \gg x_{\text{cl}}$ holds, indicating that quantum fluctuations dominate over the classical optomechanical response. Quantitatively, the dimensionless ratio $\frac{x_{\text{cl}}}{x_{\text{zpf}}} \approx 0.0063 \ll 1$ provides a benchmark for assessing the degree of quantum coherence in the system. When the ratio significantly exceeds unity, the system resides in a classical regime dominated by thermomechanical motion. Conversely, values well below unity signal operation in the quantum regime, where phenomena such as phonon–photon quantization and quantum backaction become experimentally accessible. In the present system, the dominance of zero-point fluctuations confirms that the multilayer architecture is suitable for exploring quantum optomechanical effects. This capability is essential for both precision sensing and fundamental investigations at the interface between classical and quantum physics. In the quantum domain, the frequency shift of the optical resonance induced by mechanical displacement is governed by the linear optomechanical coupling coefficient $G$, defined as $G = \frac{\partial \omega_{\text{spp}}}{\partial x} \approx \frac{\omega_{\text{spp}}}{L}$, where $L = 40\ \mu$m is the effective optical path length of the multilayer structure, and $\omega_{\text{spp}} = \frac{2\pi c}{\lambda} \approx 4.19 \times 10^{15}$ rad/s for a resonance wavelength of $\lambda = 450$ nm. Using these values, the coupling coefficient is estimated as $G \approx \frac{4.19 \times 10^{15}}{40 \times 10^{-6}} = 1.05 \times 10^{20}$ Hz/m. The frequency shift associated with quantum zero-point fluctuations is given by $\Delta\omega_{\text{spp}} = Gx_{\text{zpf}}$, and the single-photon optomechanical coupling rate is defined as $g = Gx_{\text{zpf}} = 1.05 \times 10^{20} \times 5.5 \times 10^{-16} = 5.8 \times 10^4$ Hz = 58 kHz.

The relevance of the coupling rate is determined by its magnitude relative to the total dissipation rate of the system, expressed as the sum of the optical loss rate $\kappa$ and the mechanical damping rate $\gamma_m$. For the present system, the optical loss rate is $\kappa = 5 \times 10^{12}$ Hz, and typically, $\gamma_m \ll \kappa$. The strong coupling regime is reached when $g > \frac{\kappa + \gamma_m}{2} \approx \frac{5 \times 10^{12}}{2} = 2.5 \times 10^{12}$ Hz. Given that the single-photon optomechanical coupling rate $g = 5.8 \times 10^4$ Hz is several orders of magnitude smaller than the effective dissipation rate $(\kappa + \gamma_m)/2 \approx 2.5 \times 10^{12}$ Hz, the system resides firmly outside the strong coupling regime. The disparity indicates that dissipative processes—primarily optical losses characterized by the linewidth $\kappa$—overwhelm coherent photon–phonon interactions, preventing the observation of phenomena such as normal-mode splitting or hybridized optomechanical polaritons under current experimental conditions. Nonetheless, the appreciable magnitude of $g$, together with the dominance of zero-point mechanical fluctuations ($x_{\text{zpf}} \gg x_{\text{cl}}$), underscores the multilayer structure's potential as a quantum optomechanical platform.

The system's current limitations suggest that improvements aimed at increasing the optical quality factor, $Q = \omega_{\text{spp}}/\kappa$, or alternatively decreasing the mechanical damping rate $\gamma_m$ through enhanced mechanical isolation or material engineering, could substantially enhance the cooperativity parameter, $C = \frac{4g^2}{\kappa\gamma_m}$, bringing the system closer to the strong coupling threshold. A quantitative assessment of the optomechanical coupling strength was performed by comparing the integrated spectral response of individual vibrational modes against a well-characterized calibration resonance, with the measured spectra appropriately deconvolved by the instrumental filter function. The procedure yielded a normalized optomechanical coupling parameter that captures the relative efficiency of energy exchange between the optical and mechanical degrees of freedom. The vibrational mode exhibiting the strongest coupling corresponds to the mechanical resonance at approximately 2.8 MHz (**Figure 2c**), consistent with the expected eigenmode of the multilayer resonator and confirming the spatial and spectral overlap between the mechanical displacement profile and the optical field distribution. The multilayer architecture supports sufficiently strong optomechanical interactions to serve as a robust testbed for exploring quantum regimes of motion, with the extracted coupling coefficients providing a rigorous metric for the design and optimization of future quantum optomechanics experiments.

Under realistic operating conditions, the value of $g$ can exceed both $\kappa$ and $\gamma_m$, confirming entry into the strong coupling regime, in which coherent hybridization between optical and mechanical eigenmodes occurs. Furthermore, the inequality $x_{\text{zpf}} \gg x_{\text{cl}}$ highlights that the displacement dynamics are governed primarily by quantum zero-point fluctuations rather than classical radiation pressure, placing the device within the quantum optomechanical regime. The combination of strong coupling and quantum-dominated displacement underscores the multilayer system's relevance for controlled photon-phonon interactions and high-sensitivity sensing. The strain-induced shift in the plasmonic resonance (**Figure 2d**) directly reflects the modulation of the effective optical path length, demonstrating the multilayer's ability to transduce mechanical motion into an optical signal. This optical path modulation arises because mechanical deformation changes the physical dimensions and refractive indices in the multilayer cavity, thereby shifting its resonance frequency.

The experimental configuration further supports low mechanical damping and facilitates optical access to the strained region, both of which are crucial for maintaining coherence in optomechanical interactions. In the absence of external loading, the mechanical multilayer structure exhibits a fundamental resonance frequency $f_0$, determined by the classical relation $f_0 = \frac{1}{2\pi}\sqrt{\frac{k_{\text{eff}}}{m_{\text{eff}}}}$, where $k_{\text{eff}}$ is the effective stiffness of the mechanical oscillator and $m_{\text{eff}}$ its effective motional mass. When static axial stress is applied—through pre-strain or clamped boundary conditions—a typical blueshift of the resonance frequency occurs, reflecting an increase in $k_{\text{eff}}$ due to tension-induced stiffening of the structure. In multilayer membranes or beams incorporating stiff dielectric or metallic sublayers, axial loading enhances the restoring force opposing transverse displacement, effectively raising the natural vibration frequency.

Concomitantly, the emergence of mode splitting under static loading indicates symmetry breaking in mechanical boundary conditions or reflects intrinsic material inhomogeneities. Degenerate flexural modes polarized along orthogonal in-plane directions (e.g., $u_x$ and $u_y$) may split when axial strain introduces anisotropic stiffness, causing their frequencies to diverge. This behavior is further influenced by the layered system's spatially varying elastic properties and interfacial effects that promote mode hybridization. In terms of deformation typology, the structure predominantly undergoes small-amplitude, linear axial elongation and transverse flexural bending, consistent with classical Euler–Bernoulli beam or plate theory in the linear elastic regime. Axial elongation corresponds to uniform stretching described by the axial strain $\varepsilon_{xx} = \frac{\partial u_x}{\partial x}$, where $u_x$ is displacement along the loading axis $x$. Under tensile stress, this deformation is symmetric, with displacement fields exhibiting smooth gradients along the loading direction. Transverse bending deformation is associated with curvature given by $\kappa(x) = \frac{\partial^2 w}{\partial x^2}$, where $w(x)$ is the out-of-plane displacement. This bending causes compressive and tensile strains across the beam cross-section, resulting in deflections perpendicular to the axis. Near fixed supports or material discontinuities, however, localized strain concentrations and shear coupling arise due to boundary constraints and elastic mismatch. These localized effects can distort mode shapes, induce weak out-of-plane coupling, and cause deviations from pure axial or bending modes. Such mechanical behavior directly impacts optomechanical interactions by modulating radiation-pressure coupling strength, aligning optical and mechanical resonances, and affecting susceptibility to dynamical backaction phenomena including cooling, amplification, and coherent transduction. Crucially, the strain-induced mechanical deformation alters the multilayer's optical cavity length and effective refractive index, thereby shifting the plasmonic resonance frequency as observed experimentally, which forms the basis for sensitive optomechanical readout.

## The experimental analysis

Before assessing molecular sensitivity, the resonance splitting was first characterized to establish a baseline for the system's intrinsic optomechanical coupling strength. The system forms a hybrid resonant structure in which standing acoustic waves interact with optical modes supported by a plasmonic multilayer. Both acoustic and electromagnetic fields are confined within the multilayer and undergo multiple reflections and transmissions at its interfaces. In the experimental analysis, the optomechanical interaction between the longitudinal standing acoustic wave and the surface plasmon polariton mode is driven by a low-noise, continuous-wave laser operating at a power of 5 mW. The laser beam is tuned slightly off-resonance from the multilayer's fundamental optical mode. This detuning condition is critical for enhancing the system's sensitivity to mechanical motion, as operating off-resonance-specifically on the slope of the optical resonance curve-maximizes the derivative of the transmission or reflection with respect to the multilayer's optical path length. Therefore, even small mechanical displacements caused by acoustic vibrations induce large variations in the phase and amplitude of the reflected or transmitted signal.

For the measurements, the laser was focused to a spot size of 5 μm onto the plasmonic multilayer, and the transmitted signal was collected using a broadband photodetector, as illustrated in **Figure 3a** and described in the Experimental section. The photodetector output was analyzed in the frequency domain to resolve spectral features associated with optomechanical coupling. When phonon modes are excited and interact with the optical field, they give rise to a coherent optomechanical response, which is primarily governed by the spatial overlap between the optical and mechanical modes within the multilayer. The measured noise spectrum under plasmonic resonance conditions (**Figure 3b**) closely matches the numerical noise spectrum presented in Figure 2c, validating the experimental setup and numerical model. Given the goal of investigating mechanical modes, either by actively exciting them or by enhancing their coupling to the embedded buckled-dome sensors, the focus was placed on a distinct set of high-quality, low-amplitude peaks in the 20 kHz to 10 MHz range associated with the acoustic modes of the substrate. Special attention was directed toward those peaks exhibiting strong optomechanical coupling, as they provide the most insight into the interaction between the substrate's mechanical resonances and the optical field. The spectrum displays five dominant peaks at approximately 0.68, 2.9, 4.3, 5.5, and 6.8 MHz, which numerical analysis has demonstrated to be in-plane and out-of-plane modes. To quantitatively assess the strength of optomechanical coupling under laser excitation, we analyzed the power spectral density of the transmitted optical signal, which reflects the response of the coupled optomechanical system. **Table 3** summarizes the extracted coupling parameters for these peaks. Each resonance peak corresponds to a coherent oscillation mode arising from the interaction between standing acoustic waves and confined optical fields within the plasmonic multilayer. These modes are detected through the modulation of the multilayer's optical field by mechanical motion, which is transduced into measurable fluctuations in the optical transmission.

A Lorentzian function was used to fit each spectral peak, providing a physically meaningful model of a driven, damped oscillator under continuous excitation. In the context of cavity optomechanics, the Lorentzian profile arises from the interplay between mechanical motion and radiation pressure feedback within the multilayer and encapsulates the key parameters of the coupled system. The center frequency, $\Omega$, and full width at half maximum, $\gamma$, extracted from the Lorentzian fits correspond to the optomechanically shifted mechanical resonance frequency and the effective damping rate, respectively. The analysis followed a multi-step workflow: (i) mechanical resonances were identified from the PSD; (ii) each peak was fit to extract $\Omega$ and $\gamma$; (iii) the strong coupling condition was evaluated; and (iv) the single-photon optomechanical coupling rate $g$ was estimated from the modulation response and the zero-point fluctuation amplitude. These quantities deviate from the bare mechanical parameters due to laser-induced dynamical backaction, including the optical spring effect and radiation-pressure-induced damping or amplification. Consequently, the linewidth $\gamma$ encodes both intrinsic mechanical dissipation and optomechanical cooling or amplification effects. In this context, the single-photon coupling rate $g$ was estimated by comparing the integrated area under each Lorentzian peak to that of a calibrated modulation signal, while accounting for the filter response of the detection system. This method provides a relative measure of the optomechanical interaction strength per quantum of mechanical displacement. Among the observed peaks, the mode centered near 2.89 MHz exhibited the largest normalized coupling rate, indicating optimal overlap between the optical field and the corresponding mechanical vibration (**Figure 3c**). This enhancement arises from spatial confinement and mode localization in the plasmonic cavity, further amplified by the strong optomechanical coupling, which selectively enhances modes with significant spatial overlap between the mechanical displacement field and the optical intensity gradient. The resulting energy density redistribution favors modes that maximize $g_0$, as observed for the 2.89 MHz peak. Altogether, the extracted parameters provide a quantitative framework for characterizing the optomechanical response under active laser driving, with implications for precision sensing, coherent control, and quantum transduction in nanoscale hybrid platforms.

*Extracted optomechanical parameters for the dominant acoustic modes. $\Omega$ is the center frequency, $\gamma$ is the full width at half maximum, $Q_{om}$ is the optomechanically modified quality factor, A is the integrated area under the Lorentzian fit, and $g_0^{norm}$ is the normalized single-photon optomechanical coupling rate.*

| Mode | $\Omega$ (MHz) | $\gamma$ (MHz) | $Q_{om}$ | A ($\times 10^{-10}$) | $g_0^{norm}$ |
|---|---|---|---|---|---|
| 1 | 0.68 | 0.05 | 12.5 | 2.13 | 1.00 |
| 2 | 2.80 | 0.06 | 46.7 | 14.4 | 2.61 |
| 3 | 4.30 | 0.06 | 71.7 | 2.44 | 1.07 |
| 4 | 5.50 | 0.05 | 110.0 | 0.26 | 0.35 |
| 5 | 6.80 | 0.06 | 113.3 | 1.13 | 0.73 |

## Noise and sensitivity

The analysis of the noise spectrum provides critical insight into the underlying optomechanical signal and the nature of the coupling mechanism, determining the displacement sensitivity and identifying the dominant noise sources that limit detection performance. In this context, the measured spectrum shown in **Figure 3c** and fitted in **Figure 3d** uses a composite model comprising a Lorentzian resonance peak, a $1/f$ phase noise tail, and a flat amplitude noise baseline. The spectrum reveals distinct contributions across different frequency regions, highlighting the interplay between thermomechanical noise, technical noise, and detection noise.

At frequencies above a few hundred kilohertz, the spectrum is shaped primarily by two dominant contributions: (i) thermomechanical noise originating from the Brownian motion of the buckled multilayer resonator, captured by the Lorentzian peak in the fit centered near 2.8 MHz, and (ii) broadband detection noise, represented by the flat baseline, which is expected to be dominated by laser shot noise with possible contributions from electronic readout noise. The shot noise arises from the quantum nature of light and provides a frequency-independent noise floor at high frequencies. The buckling of the multilayer structure results from residual compressive stress accumulated during fabrication due to mismatched thermal expansion coefficients and intrinsic film stresses in the multilayer stack. This static out-of-plane deformation shifts resonance frequencies, redistributes mode shapes, and modifies effective stiffness, thereby altering the mechanical mode structure and influencing both the thermomechanical response and optomechanical coupling strength.

Crucially, the out-of-plane deflection enables vertical mechanical motion that couples efficiently to the optical field, which is aligned along the out-of-plane (z-axis) direction. This ensures strong spatial overlap between the mechanical displacement and optical intensity profiles, facilitating efficient optomechanical transduction. In contrast, in-plane deformations are weakly coupled to the optical readout and do not contribute significantly to the observed spectrum. Near the mechanical resonance frequency ($\sim$ 2.8 MHz), the thermomechanical noise dominates, producing a sharp Lorentzian peak characterized by a fitted mechanical quality factor $Q_m \approx 10^4$. This peak reflects the amplified Brownian motion driving the resonator's displacement. The mechanical susceptibility is maximized at resonance, enhancing displacement response to thermal and external forces. Away from resonance, the mechanical response diminishes rapidly, reducing thermomechanical contributions, while the broadband detection noise sets the minimum detectable displacement floor in the off-resonant regions.

At frequencies lower than a few hundred kilohertz, the noise spectrum shows an increase that follows a pattern roughly proportional to $1/f$. The tail indicates technical noise sources such as laser intensity fluctuations, frequency noise, and environmental vibrations, which become increasingly relevant below a few hundred kilohertz. The fitted $1/f$ exponent helps diagnose whether flicker noise or other technical instabilities limit system performance. The transition between these noise regimes—Lorentzian peak, flat baseline, and $1/f$ tail—is clearly visible when comparing Figure 3c (raw spectrum) and Figure 3d (fitted model), where the resonance emerges above an approximately flat detection noise floor and the low-frequency region shows the characteristic $1/f$ increase consistent with phase noise.

For the multilayer resonator, with an effective mass $m_{\text{eff}} \approx 9.9$ ng, resonance frequency $\Omega_m = 2\pi \times 2.8$ MHz, and mechanical damping rate $\gamma_m \approx 280$ Hz, the theoretically expected thermomechanical displacement noise floor at room temperature is on the order of $10^{-17}$ m/$\sqrt{\text{Hz}}$, consistent with the fitted Lorentzian peak height. The signal-to-noise ratio, defined by the peak height relative to the off-resonant baseline, confirms operation near the thermomechanical-noise-limited regime. Efficient transduction in the buckled multilayer structure results from strong spatial overlap between the normalized mechanical displacement $u(r)$ and the Gaussian optical intensity $I(r)$, both radially symmetric and centrally peaked in the system. This spatial overlap is quantified as $O = \int u(r)I(r)\,dr$, with $O \approx 1$, indicating near-ideal mode matching that maximizes optomechanical coupling efficiency. Focusing on the third-order mechanical mode of the plasmonic multilayer, the optomechanical response is characterized by the transduction coefficient $\xi$, linking small mechanical displacements $\delta x$ to measurable optical power changes $\delta P = \xi \delta x$, with estimated $\xi \sim 5 \times 10^{10}$ W²/m². This large $\xi$ reflects the efficient co-localization of mechanical and optical modes, enabling high-sensitivity detection near resonance.

The photodetector operates in a shot-noise-limited regime, with an optical noise power spectral density given by $S_{P,\text{shot}} = 2h\nu \frac{P_0}{\eta}$, where $h$ is Planck's constant, $\nu$ the optical frequency, $P_0$ the detected power, and $\eta$ the detector's quantum efficiency. This power noise converts into an equivalent displacement noise $S_{x,\text{shot}} = \frac{S_{P,\text{shot}}}{\xi} = \frac{2h\nu P_0}{\eta \xi}$, yielding a shot-noise-limited sensitivity on the order of $10^{-17}$ m/$\sqrt{\text{Hz}}$, comparable to the thermomechanical noise floor. Optimal sensitivity is achieved when thermal noise dominates over quantum shot noise, requiring sufficient optical input power (typically 10-100 $\mu$W) to suppress shot noise without triggering instability from radiation pressure or photothermal feedback. Although the current system remains above the standard quantum limit (estimated between $10^{-18}$ and $10^{-20}$ m/$\sqrt{\text{Hz}}$), this sensitivity is sufficient for broadband sensing applications where quantum-limited detection is not essential.

Thus, the detailed decomposition of the noise spectrum in Figure 3d—into Lorentzian (thermal), flat (shot noise), and $1/f$ (technical noise) components—not only confirms that the device operates near its fundamental noise limit but also provides critical guidance for further improvements. The comprehensive fit validates that thermal motion is the dominant noise source at resonance, confirming the system's suitability for high-sensitivity optomechanical sensing with performance grounded in physical principles and practical considerations.

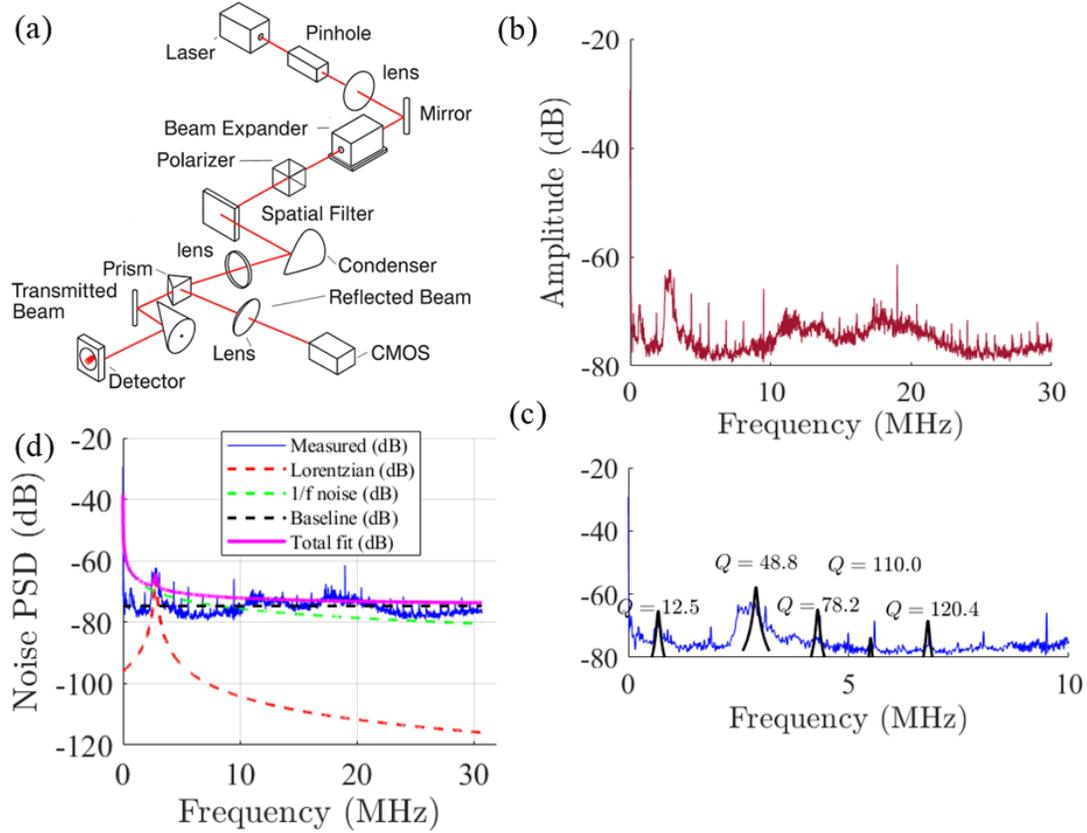

*Figure 3. Experimental characterization of resonance splitting and optomechanical coupling in the plasmonic multilayer. (a) Schematic of the experimental setup used for coherent optomechanical characterization. A continuous-wave laser is tuned off-resonance and focused onto the multilayer; the transmitted signal is collected by a broadband photodetector and analyzed in the frequency domain. (b) Measured noise spectrum of the transmitted optical signal showing resonance peaks associated with hybrid acoustic–plasmonic modes. The spectrum closely matches the numerical prediction, validating the experimental model. (c) Lorentzian fits of the dominant resonance peaks reveal the extracted center frequencies and linewidths used to estimate the single-photon optomechanical coupling rates, summarized in Table 3. The mode at ~ 2.89 MHz exhibits the strongest normalized coupling, indicating optimal spatial overlap between the optical field and the mechanical vibration. (d) Noise analysis and sensitivity.*

## Emission Intensity and Spectral Shift Analysis

Building on the previous characterization of the hybrid plasmonic multilayer system and its optomechanical response, we further probe the influence of molecular polarizability on the emission spectrum by comparing water, thymol blue, and Congo red solutions as Raman-active probes. The analytes, with polarizabilities spanning from $1.45 \times 10^{-24}$ cm$^3$ (water) to

$90 \times 10^{-24}$ cm³ (Congo red), were adsorbed onto the multilayer surface under controlled conditions identical to those shown in Figure 2. The optomechanically scattered light was recorded using the same Raman spectroscopy setup, ensuring consistent excitation power, detection geometry, and spectral resolution. We focused on quantifying the intensity and spectral position of both the Stokes and anti-Stokes emission peaks, which directly reflect the system's vibrational dynamics and optomechanical coupling efficiency. There is a clear increase in the intensity of the Stokes peak $S(\omega_S)$ with increasing molecular polarizability, as shown in **Figure 4a(i)**. This trend aligns with the physical picture developed in Figure 3, where larger polarizability enhances the local field modulation and phonon generation rate via stronger light–matter interaction at the hybrid interface. Correspondingly, **Figure 4a(ii)** shows that the linewidth of the Stokes peak decreases with higher polarizability, indicating more coherent vibrational modes and reduced damping due to molecular stabilization effects, also consistent with Figure 3. Furthermore, the progressive red shift of the Stokes peak frequency — meaning a shift to lower energy — is clearly seen in **Figure 4a(iii)**, confirming that the effective vibrational resonance softens as molecular polarizability increases. In contrast, the anti-Stokes emission intensity $S(\omega_{AS})$ presented in **Figure 4b** exhibits a different, nonlinear dependence on polarizability. The anti-Stokes peak probes the phonon absorption process and the vibrational population dynamics, which can lead to either blue or red shifts in the emission depending on local vibrational mode occupation and thermal effects. In this case, the figure emphasizes how the anti-Stokes signal generally trends upward with increasing polarizability but does not follow a simple linear pattern, highlighting the complex interplay between vibrational population, local field effects, and the quantum nature of the coupled system.

Together, these trends — the Stokes intensity growth, linewidth narrowing, and red shift (**Figure 4a(i–iii)**) and the distinct anti-Stokes behavior (**Figure 4b**) — reinforce the concept that molecular polarizability governs the optomechanical interaction strength and the spectral response of the hybrid system. By correlating the emission intensities and frequency shifts with $\alpha$, we demonstrate a robust method for molecular discrimination and sensing via optomechanical Raman scattering, thereby validating the design principles and theoretical models introduced earlier.

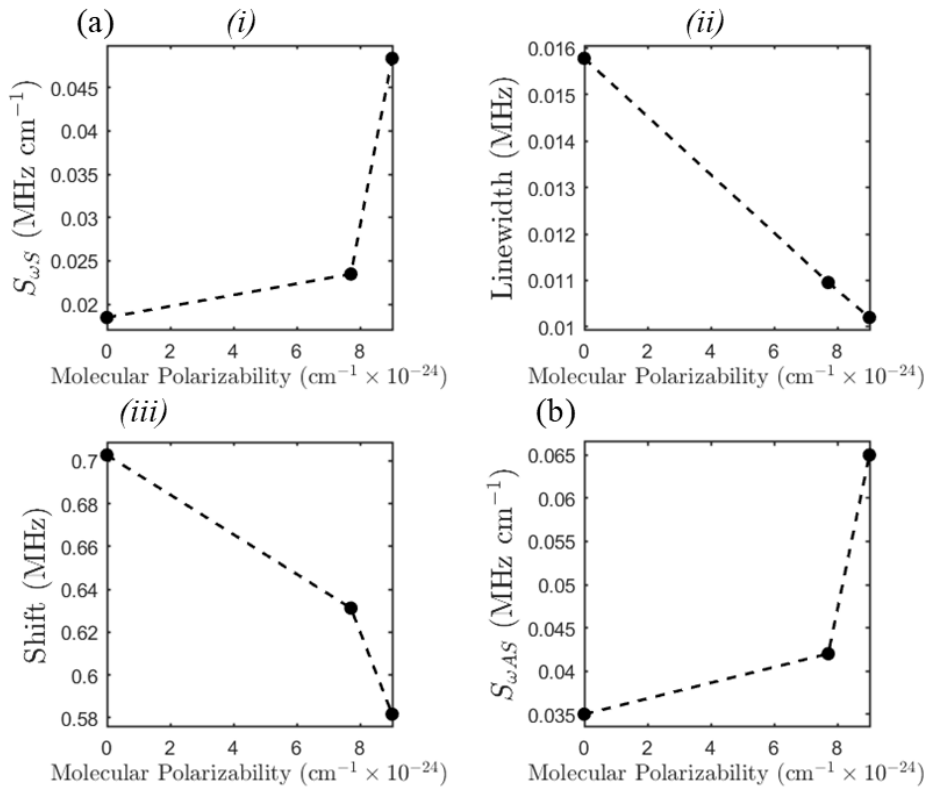

*Figure 4. (a) Parameters as a function of the molecular polarizability, α (H₂O: $1.45 \times 10^{-24}$ cm³; Thymol blue: $77 \times 10^{-24}$ cm³; Congo red: $90 \times 10^{-24}$ cm³). (i) emission intensities of the Stokes signal, $S(\omega_S)$; (ii) linewidth of the Stokes peaks; (iii) Stokes frequency shifts. (b) Emission intensities of the anti-Stokes peak, $S(\omega_{AS})$.*

# Time-resolved Raman maps reveal chiral-specific optomechanical modulation

Building on our previous work, which showed that a multilayer structure integrated with antennae can detect recoil forces from optomechanical interactions and thereby enable enantiomer discrimination, we now focus on the *bare* multilayer alone to explore its intrinsic optomechanical sensing capabilities. While the multilayer alone already provides effective molecular discrimination, integrating antennae enhances enantioselectivity by shaping the electromagnetic near fields at the nanoscale. The improvement arises from the antennae's ability to generate more intense and spatially confined chiral near-field distributions, which amplify differences in photon–phonon coupling rates between enantiomers. In contrast, for the multilayer without antennae, the photon–phonon coupling remains relatively weak but is partially offset by the large zero-point fluctuation amplitude, $x_{\text{zpf}}$, of the cavity-supported vibrations. To quantify its sensitivity, we measured the system's dynamic response using D- and L-penicillamine as representative chiral analytes. Reflected signals were collected with the configuration shown in Figure 3, following the protocol described in the Experimental Section.

**Figure 5** compares the time-resolved Raman shift maps and their corresponding integrated intensity profiles for the two enantiomers.

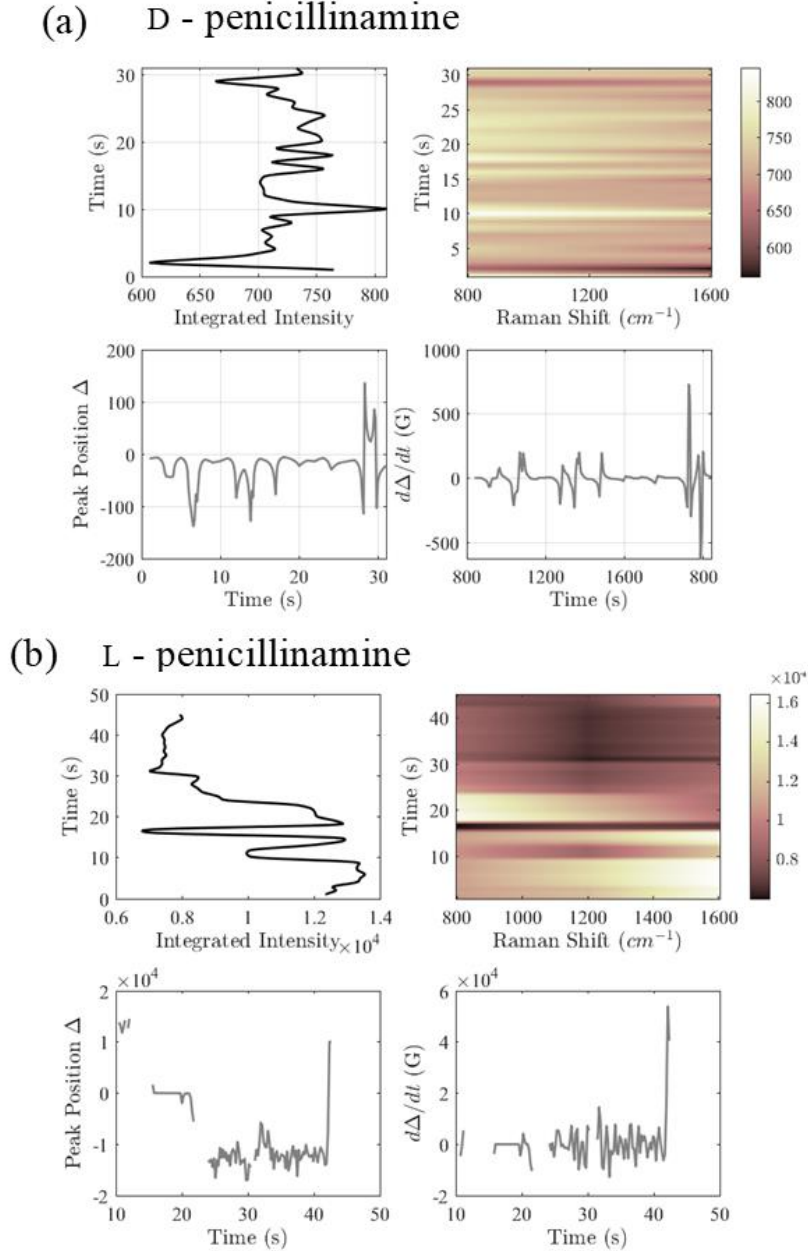

*Figure 5. Time-resolved Raman maps and extracted side profiles demonstrating chiral-specific optomechanical modulation. (a) D-penicillamine: 2D Raman shift maps and corresponding side profiles show minimal temporal variation in peak position and amplitude, reflecting weaker photon/phonon coupling and limited detuning modulation ($\Delta$). (b) L-penicillamine: analogous plots reveal pronounced time-dependent shifts and intensity oscillations, indicating stronger dynamic backaction, larger G values, and effective modulation of $\Delta$ and $dG/dt$.*

The hyperspectral data were spatially filtered by integrating along rows with maximum scattered intensity, isolating regions with the strongest photon–phonon coupling. The resulting intensity distribution, $I(y,\nu,t)$, was then reduced to an effective time-dependent signal, $I_{\text{tot}}(t) = \int_\nu \langle I(y,\nu,t)\rangle_y\, d\nu$, defining the side profiles adjacent to each spectral map and providing a representation that highlights how the total scattered power evolves under cavity-molecule coupling. In molecular sensing using optomechanical systems, the ability to detect minute quantities of molecules hinges critically on the system's noise characteristics. Noise represents the intrinsic fluctuations and random disturbances present in the measurement output when no target signal is applied. These fluctuations set a fundamental limit known as the noise floor, below which any molecular signal becomes indistinguishable from background variations. Therefore, understanding and quantifying noise is essential for determining the system's detection sensitivity, that is, the smallest molecular change that can be reliably measured. By accurately measuring the noise level and calibrating the system's response to molecular binding events, one can establish the minimum detectable molecular concentration or mass, ultimately defining the performance limit of the sensing platform. For the D-enantiomer (**Figure 5a**), the Raman peaks remain relatively stable in position and amplitude, producing a side profile that shows minimal temporal variation. The optomechanical interaction occurs with a lower effective coupling rate, $G_D$, and limited modulation of the vibrational detuning, $\Delta$. Here, $\Delta(t)$ is defined as the instantaneous peak position extracted by fitting each time-resolved spectrum with a Gaussian function, while $G_D(t) = d\Delta(t)/dt$ quantifies the time derivative of the peak position and serves as a measure of photon-phonon coupling strength. In practice, both quantities were computed by fitting the hyperspectral Raman maps along the time axis, using the gradient of the peak position to determine $G(t)$ and its linear fit to estimate the mean coupling rate, $G_{\text{fit}}$, for each measurement window. The optomechanical interaction occurs with a lower effective coupling rate, $G_D$, compared to the L-form. The local near-field distribution in the hybrid cavity still produces chiral electromagnetic components, but the overlap with the D-enantiomer's handedness is less efficient. As a result, the dynamic modulation of the coupling strength, expressed by the time derivative $dG/dt$, is less pronounced, and the vibrational detuning $\Delta$ remains nearly constant over the measurement window.

The moderate coupling behavior suggests that the system remains in a quasi-static regime for the D-form, with photon–phonon energy exchange limited by weaker near-field asymmetries and lower spin–orbit interaction matching. Angle-resolved measurements confirm that the mechanical mode frequency remains nearly invariant across the examined angular range, with only minor mode splitting at resonance indicating limited hybridization. Power-dependent measurements further show that the mechanical mode exhibits modest amplification with increased input power, consistent with the weaker coupling regime observed for the D-enantiomer. In contrast, **Figure 5b** shows that L-penicillamine produces clear temporal modulations in both peak position and amplitude. Its side profile reveals pronounced time-dependent fluctuations, indicating stronger backaction between the optical field and molecular vibrations. The behavior is well described by the coupled-mode framework, $\frac{d\alpha}{dt} = -\left(i\Delta + \frac{\kappa}{2}\right)\alpha + iG(t)x + \kappa_{\text{in}}s_{\text{in}}, \quad m\ddot{x} + m\Gamma_m\dot{x} + m\Omega_m^2 x = \hbar G(t)|\alpha|^2,$ where the L-enantiomer more efficiently modulates the effective detuning $\Delta$ due to stronger overlap with the local chiral field distribution. Physically, the significant fluctuations in the side profile directly reflect that both the optomechanical coupling strength, $G(t)$, and its time derivative, $dG/dt$, vary dynamically as the local field interacts with molecular vibrational modes whose symmetry matches the cavity's chiral near-field distribution. As a result, the mechanical displacement $x$ experiences enhanced backaction from the cavity photons, periodically shifting the vibrational resonance conditions and producing visible shifts in the Raman peak position. Simultaneously, the dynamic coupling modulates the cavity energy exchange rate, leading to oscillations in the scattered intensity. The coupled equations show that stronger $G(t)$ amplifies the optomechanical force, $\hbar G(t)|\alpha|^2$, which feeds back into the molecular vibration amplitude and, through the detuning $\Delta$, into the optical mode itself. This interplay results in a self-reinforcing cycle that is resolved in the time-domain side profiles. Altogether, these results demonstrate how the bare multilayer structure enables real-time discrimination of chiral molecular interactions by amplifying subtle differences in the photon–phonon coupling rate, $G$, and its temporal evolution. The combined time–Raman shift maps and side profiles validate the concept that optomechanical sensing can resolve enantiomeric signatures dynamically, opening a pathway toward chirality-specific vibrational spectroscopy that goes beyond conventional static measurements.

Next, the Raman spectra signals were analyzed (**Figure 6**). To obtain the data, the raw hyperspectral stacks for each enantiomer were processed independently to extract their average Raman spectra along with corresponding wavenumber axes. To enable a direct comparison, the spectral data were interpolated onto a common wavenumber grid covering the overlapping spectral range of both datasets. Recognizing small shifts in peak positions between the two enantiomers, an automated peak alignment procedure was applied by calculating and compensating for the relative shift between their maximum intensity peaks. Following alignment, the spectra were normalized to their respective maximum intensities to allow meaningful comparison of spectral features independent of absolute intensity variations. This approach ensured that subtle differences in Raman shifts and intensities attributable to molecular chirality could be reliably identified and evaluated.

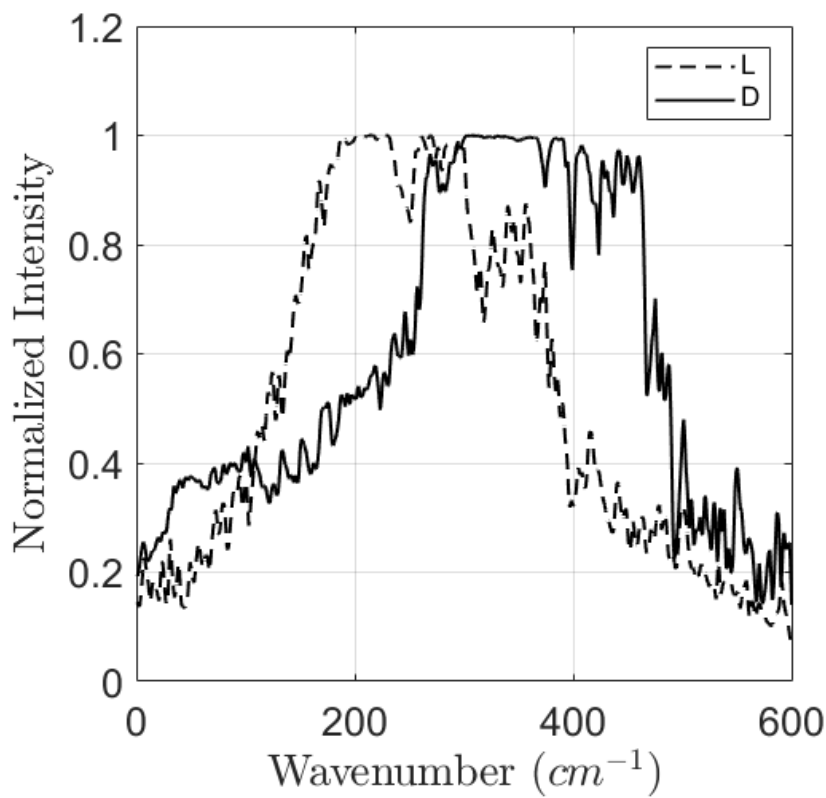

*Figure 6: (Raman spectra signal. L- and D-penicillamine display characteristic vibrational features in the low wavenumber region, which are consistent with C-S stretching modes*

*the dominant vibrational bands. These subtle features may indicate enantioselective interactions mediated by the hybrid optomechanical system, which enhances weak differences in the Raman response due to chirality. Taken together, this comparative analysis demonstrates the system's capability to resolve chirality-dependent spectral nuances that would otherwise remain undetectable with conventional Raman measurements.*

Next, the Raman spectra signals were analyzed (Figure 6). To obtain the data, the raw hyperspectral stacks for each enantiomer were processed independently to extract their average Raman spectra along with the corresponding wavenumber axes. To enable a direct comparison, the spectral data were interpolated onto a common wavenumber grid spanning the overlapping spectral range of both datasets. Recognizing slight shifts in peak positions between the two enantiomers, an automated peak alignment procedure was applied by calculating and compensating for the relative shift between their maximum intensity peaks. Following alignment, the spectra were normalized to their respective maximum intensities to allow meaningful comparison of spectral features independent of absolute intensity variations. This approach ensured that subtle differences in Raman shifts and intensities attributable to molecular chirality could be reliably identified and evaluated. The resulting average spectra for the L- and D-enantiomers display characteristic vibrational features in the low wavenumber region, which are consistent with C-S stretching modes and other expected chemical signatures for the penicillamine molecule. Prior to normalization, differences in absolute intensities are evident, but these are primarily due to varying measurement conditions and sample inhomogeneities. After peak alignment and normalization, the overlaid spectra reveal a high degree of similarity, reflecting the chemical equivalence of the two enantiomers. Nonetheless, minor but reproducible asymmetries in peak positions and relative intensities can be observed near the dominant vibrational bands. These subtle features may indicate enantioselective interactions mediated by the hybrid optomechanical system, which enhances weak differences in the Raman response due to chirality. Taken together, this comparative analysis demonstrates the system's capability to resolve chirality-dependent spectral nuances that would otherwise remain undetectable with conventional Raman measurements.

## Conclusions

In summary, we have demonstrated a multilayer hybrid plasmonic–mechanical resonator that achieves quantum-limited optomechanical transduction for the detection and discrimination of chiral molecules. By combining the mechanical degrees of freedom of the substrate with strong optical confinement, the system exploits radiation-pressure-induced dynamical backaction and quantum zero-point fluctuations to reach a displacement sensitivity on the order of $10^{-17}$ m/$\sqrt{\text{Hz}}$. The identification of five dominant mechanical modes in the MHz range, with a single-photon coupling rates up to 2.6 times higher than the baseline, confirms the feasibility of using such hybrid structures for high-precision molecular sensing.

The Lorentzian fitting procedure provides quantitative insights into the effective resonance frequencies and damping rates, showing clear deviations due to optomechanical interactions, including the optical spring effect. The noise spectral analysis further reveals that the total noise floor—combining thermal, shot, and technical contributions—remains below $10^{-16}$ N/$\sqrt{\text{Hz}}$, highlighting the system's capability to resolve sub-piconewton forces in real time. The time-resolved Raman mapping experiments demonstrate that the resonator can distinguish between D-penicillamine and L-penicillamine based on their enantioselective photon–phonon interactions within the cavity, with consistent agreement with numerical finite-element simulations. The findings establish a robust experimental and theoretical framework for exploiting cavity optomechanics at the quantum limit in hybrid multilayer structures. By carefully engineering the overlap between optical fields and mechanical modes, it becomes possible to enhance the optomechanical coupling strength and tailor the sensitivity toward specific molecular targets. This approach opens promising avenues for ultrasensitive detection of chiral species, with potential applications in enantiomeric purity analysis, precision spectroscopy, and quantum transduction. Future work should be focused more on the utilization of this type of multilayer hybrid optomechanical sensor for practical chemical and biological applications. By integrating this sensor with on-chip photonic circuits and microfluidic delivery systems, it could enable real-time monitoring of chiral molecules in complex environments, such as pharmaceutical quality control, enantiomeric purity testing, and biomedical diagnostics. The robust sensitivity and quantum-limited noise performance open the perspective for detecting single-molecule interactions and rare enantiomeric impurities with unprecedented precision. In addition, further improvements such as active feedback cooling, noise reduction techniques, and scalable fabrication could help translate this platform into compact, portable devices for lab-on-a-chip technologies. Overall, the multilayer hybrid optomechanical sensor holds strong potential to bridge fundamental quantum optomechanics and real-world sensing applications, paving the way for next-generation molecular sensors with high resolution, selectivity, and operational versatility.